\documentclass[aip,graphicx]{revtex4-1}
\usepackage{graphicx} 
\usepackage{multirow}
\usepackage{xcolor}

\draft 

\begin{document}

\title{Nanoparticles of NbC produced by laser ablation in liquid: a study of structural, magnetic and superconductivity properties.} 



\author{Fernando Fabris}
\email[]{ffabris@ifi.unicamp.br}
\affiliation{Instituto de F\'{i}sica Gleb Wataghin, UNICAMP, 13083-859 Campinas, S\~{a}o Paulo, Brazil}

\author{Ali F. Garc\'{i}a-Flores}
\affiliation{Instituto de F\'{i}sica Gleb Wataghin, UNICAMP, 13083-859 Campinas, S\~{a}o Paulo, Brazil}

\author{Julian Andres Munevar Cagigas}
\affiliation{Centro de Ciências Naturais e Humanas, UFABC, S\~{a}o Bernardo do Campo, S\~{a}o Paulo, Brazil}

\author{Jos\'{e} Javier S\'{a}ez Acu\~{n}a}
\affiliation{Centro de Ciências Naturais e Humanas, UFABC, S\~{a}o Bernardo do Campo, S\~{a}o Paulo, Brazil}

\author{Carlos Rettori}
\affiliation{Instituto de F\'{i}sica Gleb Wataghin, UNICAMP, 13083-859 Campinas, S\~{a}o Paulo, Brazil}

\author{Ricardo R. Urbano}
\affiliation{Instituto de F\'{i}sica Gleb Wataghin, UNICAMP, 13083-859 Campinas, S\~{a}o Paulo, Brazil}


\date{\today}

\begin{abstract}

Niobium carbide (NbC) is a high-field Type II superconductor with a critical temperature ($T_C$) of 11.1 K, just above that of pure Nb ($T_C = 9$ K). Downsizing NbC to the nanoparticle scale introduces significant alterations in its critical field and/or the superconducting temperature. Here we report on superconducting NbC nanoparticles with $T_C \approx$ 10 K synthesized by laser ablation in acetone, using the lens-target distance (laser fluence) and centrifugation as control parameters of the particle size. X-ray diffraction analyses certified the cubic NbC phase and electron microscopy images revealed spherical particles with average size near 8 nm, with no apparent size dependence on fluence. Besides, magnetization curves exhibited magnetic loops featuring a saturation magnetization around $10^{-3} \mu_B$/molecule along with a small and typical superconducting loop for all investigated samples. We also observed a suppression of the diamagnetic behavior below $T_C$ upon decreasing laser fluence. Moreover, all samples exhibited a weak electron spin resonance (ESR) Curie-like signal at $g\approx2.0$ probably associated with localized defects in the particle's surface. The intriguing coexistence of superconductivity and magnetism in nanoparticles has recently garnered significant research attention. This complex scenario and unique properties are due to the substantial increase of surface-to-volume ratio in these superconducting NbC nanoparticles and further investigation would be crucial to unveil novel material properties and shed new light on our understanding of the superconducting phenomenon in this new morphology.

\end{abstract}
\pacs{}
\maketitle

\section{Introduction}

Niobium carbide (NbC), a representative transition metal carbide, possesses distinctive physical properties featuring a simple NaCl-type cubic lattice, an elevated melting point ($3600^{\circ}$C), remarkable hardness, high thermal and electrical conductivity, superior wear resistance, and chemical stability\cite{Cuppari2016}. Such a versatile combination renders NbC suitable for various applications, including cutting and grinding tools\cite{UHLMANN2019, MONTENEGRO2018}, wear protection of surfaces\cite{WOYDT2016, WOYDT2013}, catalysis\cite{Alhowity2024}, super-capacitors and hybrid ion capacitors\cite{Yang2021}. As a Type II superconductor, NbC is considered a key player in high-field superconductor applications, boasting excellent critical superconducting properties. Notably, within the realm of low-temperature superconductors, NbC stands out with a relatively high superconducting transition temperature of $T_C=11.1$ K, just above that of pure Nb ($T_C=9$ K), coupled with an impressive critical magnetic field ($H_{C2}$) exceeding 4 T\cite{Kaustav2016,zou2011,Yingying2012}.

Despite extensive research since its discovery in 1964\cite{Wells1964}, considerable attention has been devoted to understand the superconducting properties of NbC. Investigations have explored alterations in carbon vacancy density, varied production routes, and diverse synthesis techniques\cite{Gusev1989}. Additionally, numerous studies have probed the superconducting properties of NbC in low-dimensional forms, such as thin films, nanofibers, and nanoparticles, exhibiting different morphologies\cite{Li2006,Fukunaga1998,Kaustav2016}. However, challenges persist in the production of bulk or nanosystems based on NbC with specific functionalities, primarily due to the high temperatures required. Nevertheless, the physical top-down synthesis method namely pulsed laser ablation have demonstrated capability of producing NbC nanomaterials and it stands out as a ``green" technique enabling the preparation of a substantial quantity of materials at room temperature\cite{Zhang2019,ZHANG2016,Amendola2013,Frias2022}. This process involves focusing a high-power pulsed laser on the Nb target surface, leading to the production of NbC nanoparticles in organic solvents (such as acetone or ethanol). Interestingly, these particles are surrounded by a thin porous graphite layer, enhancing the chemical stability of the solution.\cite{Zhang2019}

Regardless of the substantial body of research on niobium carbide, there is a notable gap in data exploring the coexistence of magnetic and superconducting properties of NbC nanoparticles produced through laser ablation. Therefore, in this work, we dedicated our efforts to the production and a detailed characterization of NbC using laser ablation in acetone. By varying the laser fluence at the Nb target surface, we examined changes in the morphological, magnetic, and superconducting properties. A comparative analysis between melt particles (with average diameter about 470 nm) and particles formed through target vaporization (with average diameter about 8 nm) provided new insights into crucial physical properties of NbC nanoparticles. The magnetic properties of NbC nanoparticles manifested in three components: an electron spin resonance (ESR) signal characterized by a Curie-like $T$-dependent behavior attributed to a localized ESR active defects confined within the surface of the NbC nanoparticles; a paramagnetic component observed in all samples and all temperatures, linked to typical surface effects in materials reduced to the nanometric scale, and finally, the superconducting behavior marked by a sudden diamagnetic behavior below a critical temperature $T_C\approx$10 K for all studied samples. We provide a quantitative analysis of the superconducting volume and critical fields for different laser fluences used in the nanoparticles production process, shedding new light on the intricate interplay between diamagnetic and superconducting properties of NbC nanoparticles.

\section{Material and methods}

The NbC nanoparticles were synthesized through pulsed laser ablation (LA) using a Nb target immersed in acetone. The LA process employed a Nd:Yag laser (1064 nm) with a 10 ns pulse duration and a 20 Hz repetition rate. Focused with a 50 mm lens, the laser was directed onto a freshly polished (sanded) surface of a Nb target ($99.9\%$ of purity) vertically positioned and submerged in 75 ml of acetone. The laser path through the liquid spanned a distance of 2 cm.


The ablation process continued for 20 minutes, resulting in a uniformly dark and stable solution. Initially, the NbC nanoparticles were synthesized under a laser fluency of 20 J/cm$^2$. The resultant sample underwent centrifugation at 1000 RCF for 10 minutes, separating the supernatant (S1) from the sediment (P1). Subsequently, three additional samples were produced with laser fluencies of 45 J/cm$^2$, 70 J/cm$^2$, and 100 J/cm$^2$, each of them subjected to the same procedure aforementioned. The resulting supernatant samples were named S2, S3, and S4, respectively. Finally, the powder samples investigated here were obtained from evaporating the acetone at room temperature.

Total reflection X-ray fluorescence (TXRF) elemental analysis of the samples revealed an exceptional purity level of 99.5 \% following the entire experimental procedure. The remaining 0.5 \% primarily comprises iron contaminants, presumably coming from the sanding paper. 


The X-ray diffraction (XRD) measurements of the samples were performed in a Phaser D2 diffractometer (Bruker). Transmission electron microscopy (TEM) images were taken in a Talos F200X G2 equipment equipped with a field emission gun (FEG-X) source operating at 200 kV and in JEOL JEM-2100F also with a FEG gun source operating at 200 kV. TEM samples were prepared by dropping the acetone solution containing the nanoparticles on a copper grid with an ultra-thin hollow carbon film. The \textit{T}-dependent \textit{dc}-magnetic susceptibility measurements of the powdered samples were carried out in a SQUID (Quantum Design, MPMS 3) magnetometer. The ESR experiments were carried out in a Bruker ELEXYS 500 ESR spectrometer at $\sim 9.5$ GHz (X-band), using an appropriate TE102 resonator coupled to an Oxford \textit{T}-controller helium gas flow system varying the temperature in the range of $4.2$ K $< T < 300$ K. Colloidal NPs and powdered bulk samples were dispersed in high purity paraffin for the ESR experiments. 

\section{Results and Discussion}

\begin{figure}
	\centering
	\includegraphics[width=0.5\textwidth]{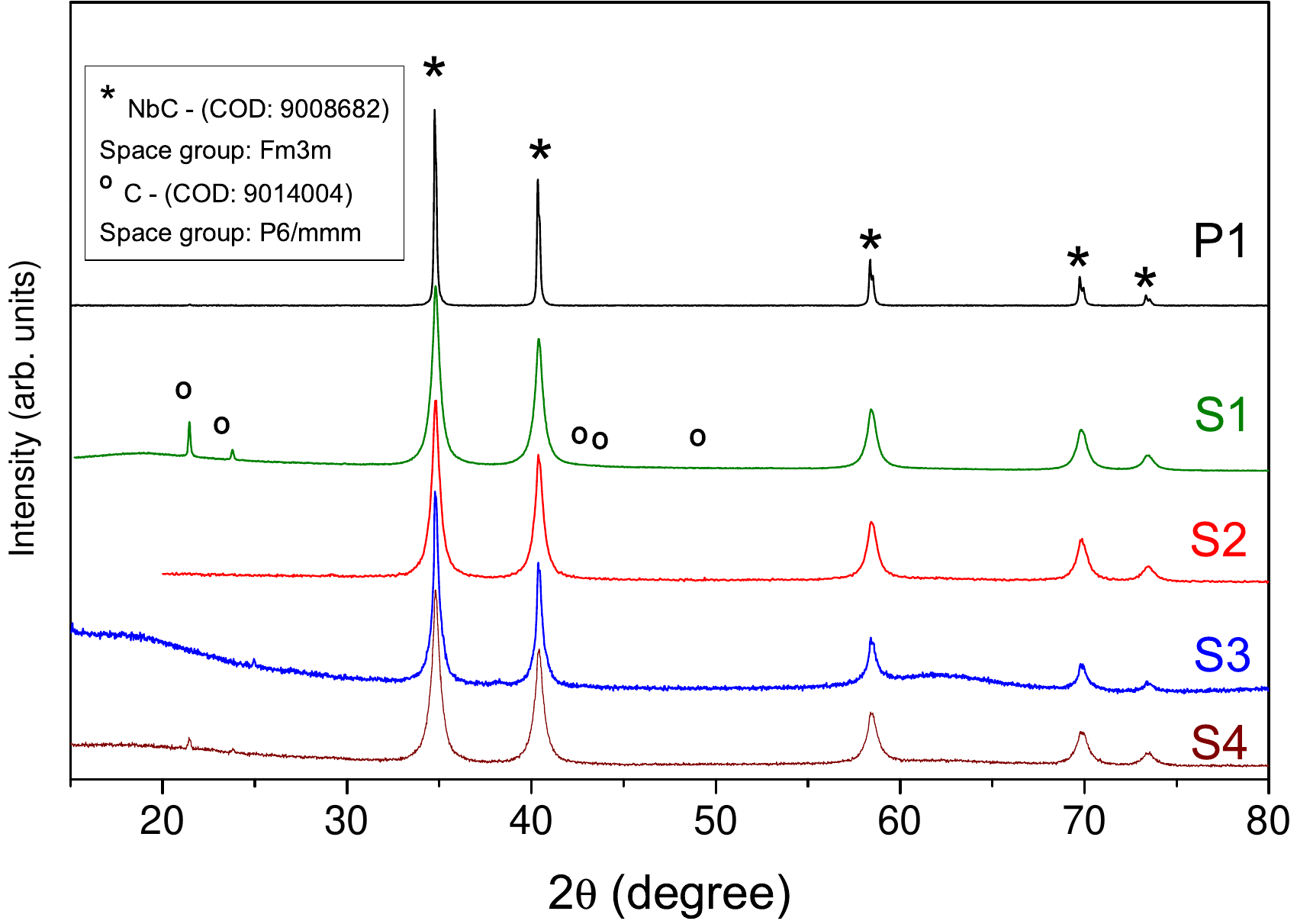}
	\caption{\label{DRX} XRD patterns of samples P1, S1, S2, S3 and S4. The stars on the figure represent the cubic NbC diffraction pattern (COD: 9008682) and the circles the hexagonal graphite pattern (COD: 9014004).}
\end{figure}

Figure \ref{DRX} illustrates the X-ray diffraction (XRD) patterns for the samples. All discernible diffraction patterns agree well with a cubic NbC structure (COD database code: 1011323). The diffraction peaks exhibit notable broadening beyond the instrumental resolution, indicative of a finite-size effect. Notably, the peak broadening observed in the P1 sample is less pronounced compared to the Si samples, suggesting a larger crystal size for P1. 
The determined lattice parameter for all nanoparticles samples is $a=4.469(1)$~\AA, in close proximity to the expected value for stoichiometric NbC ($a=4.470$ $\AA$)\cite{Cuppari2016}. This suggests a negligible presence of carbon vacancies in the nanoparticles generated through laser ablation.\cite{Cuppari2016} Additionally, certain samples exhibit peaks associated with hexagonal graphite (COD database code: 9014004).

\begin{figure}
	\centering
	\includegraphics[width=\textwidth]{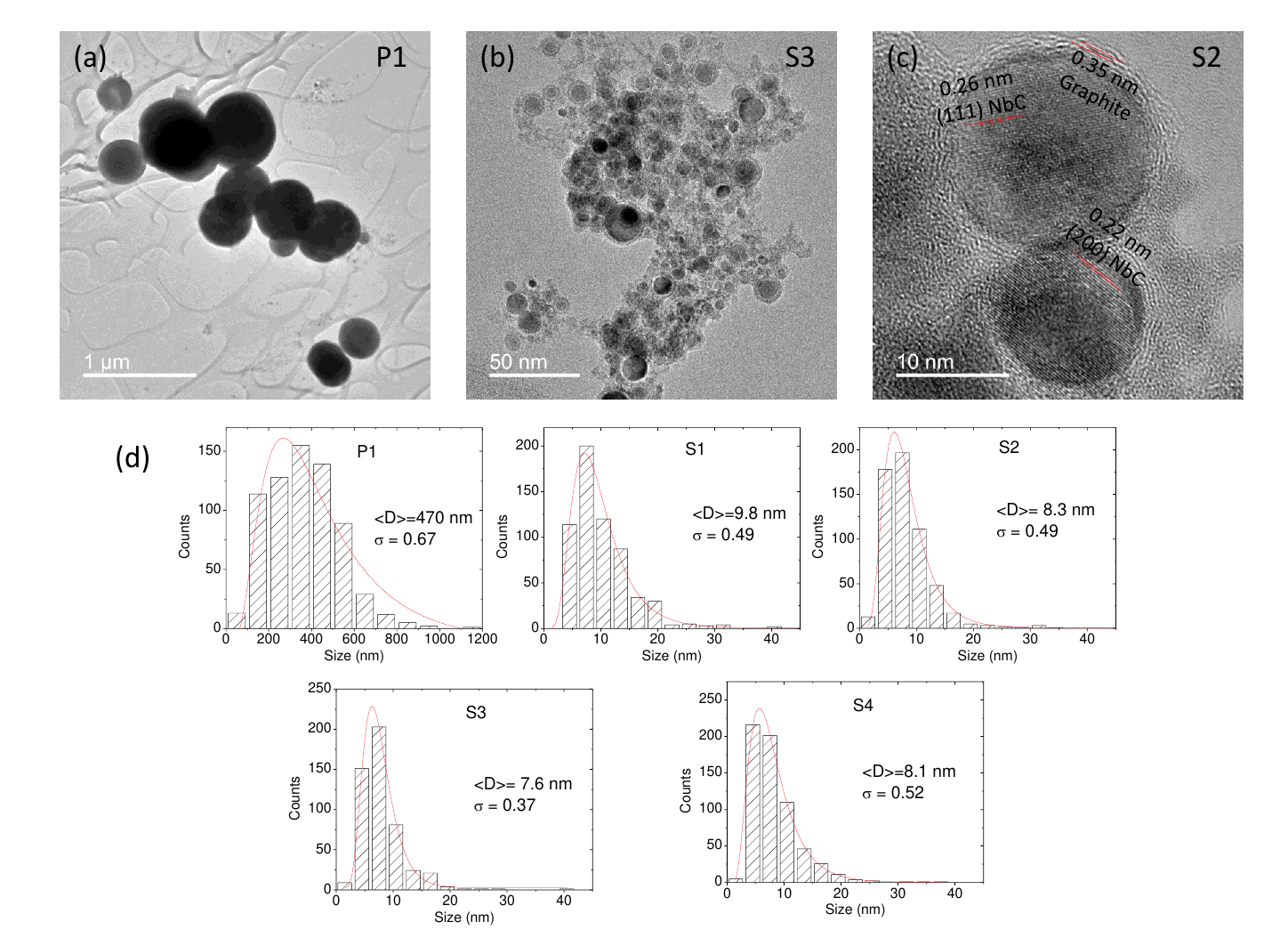}
	\caption{\label{TEM} Representative TEM images of P1 (a), S3 (b) and S2 (c) samples. The sample size distributions are presented in (d). The red lines are the best fit with a log-normal: the mean diameter ($\langle D\rangle$) and the standard deviation ($\sigma$) are displayed.}
\end{figure}

Figure \ref{TEM} displays representative transmission electron microscopy (TEM) images of P1, S2, and S3 samples, all exhibiting a similar spherical morphology with a broad size distribution, which was determined by counting nearly thousand particles in each specimen, as shown in Figure \ref{TEM} (d). The associated size histograms were fitted with a log-normal function defined by $f(D)=(\sqrt{2\pi}\sigma_0 D)^{-1}\exp\left[-\frac{\ln^2(D/D_0)}{2\sigma_0^2}\right]$ and, thus, $D_0$ and $\sigma_0$ were estimated. Therefore, the mean diameter $\left\langle D\right\rangle=D_0\exp\left(\frac{\sigma_0^2}{2}\right)$ and standard deviation $\sigma=\left\langle D\right\rangle\left[\exp\left(\sigma_0^2\right)-1\right]^{1/2}$ were calculated.

It is noteworthy that the sedimented sample particles are considerably larger than those in the supernatant. In the former, the ablation process is broadly categorized as fragmentation, involving the detachment of hot atoms, vapors, and liquid drops out of thermodynamic equilibrium conditions\cite{Amendola2013}. Specifically, the P1 particles are formed by the melt expulsion of the target during the ablation process. Conversely, in the supernatant (Si) samples, the nanoparticles are generated after the detachment of hot atoms and vapors. Notably, the $\left\langle D\right\rangle$ and $\sigma$ for Si samples do not substantially change with fluency and all of them exhibited nearly 8.5(5) nm size.

High-resolution TEM images reveal crystalline planes of NbC, as illustrated in Figure \ref{TEM}(c) for the (111) and (200) planes in two different particles. Additionally, a carbon shell is evident on the particle surface. Clear lattice fringes with an inter-planar spacing of 0.34 nm align precisely with the typical (002) plane of graphite carbon. In all S samples, 3 to 6 layers of graphite surround the NbC nanoparticles, whereas for the P1 sample, 9 to 20 carbon layers cap the NbC particles. Similar carbon encapsulated metals are obtained by the ablation of Ti, V, Nb, Cr, Mo, W, Zr, Ni and Co\cite{Zhang2019,ZHANG2016,ZHANG2013,Debonis2020}.

\begin{figure}
	\centering
	\includegraphics[width=0.5\textwidth]{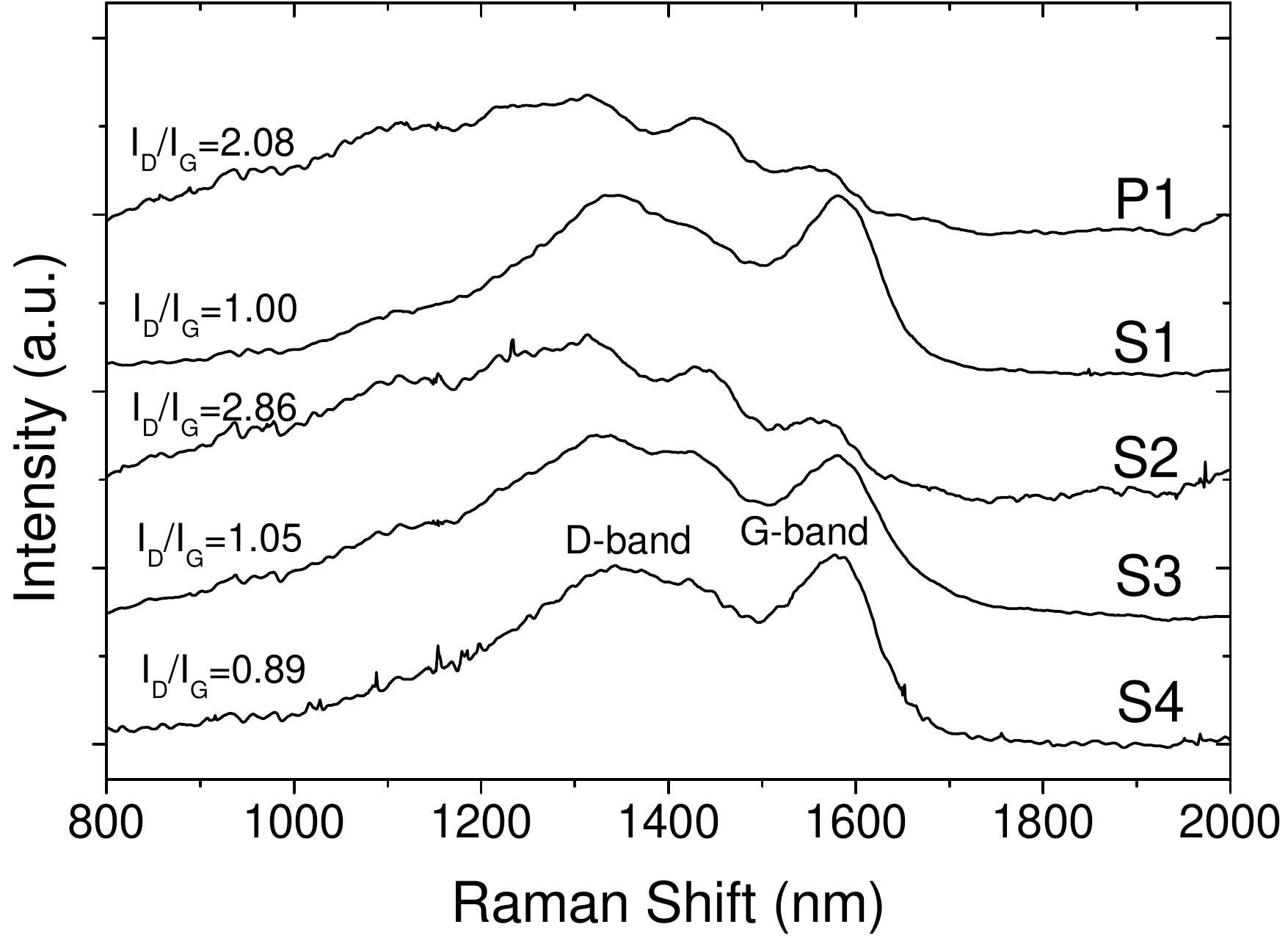}
	\caption{\label{Raman}Room temperature Raman spectra for all carbon-encapsulated NbC samples.}
\end{figure}

Raman spectroscopy was employed to validate the presence of carbon and its crystalline structure in the carbon-encapsulated NbC samples. The Raman spectra of all five samples, as depicted in Figure \ref{Raman}, reveal two prominent bands at approximately 1350 cm$^{-1}$ (D band) and 1600 cm$^{-1}$ (G band). These bands correspond to the vibrations of carbon atoms with dangling bonds in the in-plane terminations of disordered graphite (D) and the vibrations in sp$^2$-bonded carbon atoms in a 2D hexagonal lattice (G), respectively, as established by Emanich et al. \cite{Nemanich1979}. The relative intensity ratio $I_D/I_G$ varies across the sample regions, ranging from 0.89 to 2.89 for all samples. Although a precise value for $I_D/I_G$ cannot be unequivocally determined, the consistent elevated values suggest the presence of numerous structural defects in the synthesized carbon shells. This observation is in accordance with other carbon-encapsulated nanoparticles produced by laser ablation\cite{Zhang2019, ZHANG2016}.

\begin{figure}
	\centering
	\includegraphics[width=0.5\textwidth]{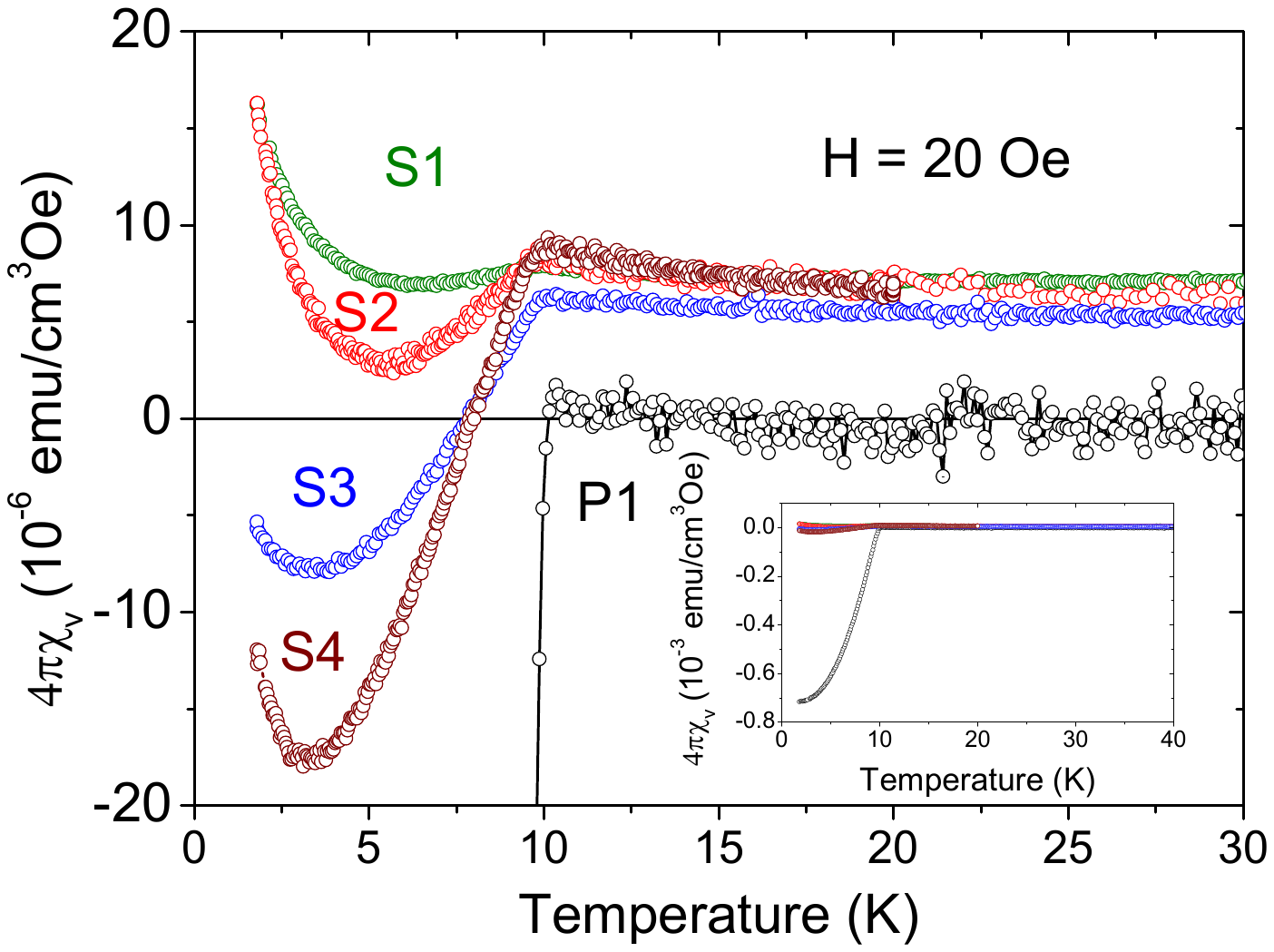}
	\caption{\label{Mag-MT}Zero field cooled magnetization at 20 Oe for all samples. Inset displays the zooming out of these measurements.}
\end{figure}

Magnetic measurements are presented in Figure \ref{Mag-MT} (a). The zero-field cooled (ZFC) magnetization as function of temperature revealed the onset of the superconducting critical temperature at $T_C \simeq$ 10 K. This value is slightly smaller than that of stoichiometric bulk NbC ($T_C=11.1$ K)\cite{Giorgi1962}, but aligns with reported $T_C$ values for NbC nanoparticles obtained through other methods\cite{Kaustav2016}. Besides, no dependence of $T_C$ with the laser fluency employed in the ablation process was observed.

Moreover, a positive magnetization was observed for the S samples above $T_C$, attributed to a paramagnetic contribution of C defects on the NbC surface nanoparticles. Interestingly, the magnetization was nearly zero for the P1 sample, within the accuracy of the equipment.

Below the critical temperature ($T_C$), a notable decrease in the superconducting volume was observed when comparing the P1 sample with the S samples. Below $H_{C1}$ in a bulk type II superconductor, the magnetic field is completely expelled from the sample volume, except for a surface layer with a thickness denoted by $\lambda_L$ known as the London penetration length. Under ideal conditions, the magnetic susceptibility below $T_C$ is expected to be $\chi_{bulk}^{SC}=-1/4\pi$. For bulk NbC, $\lambda_L$ is determined as 140 nm\cite{Shang2020}, meaning that the magnetic field only penetrates into the P1 particles partially, whereas it penetrates into the S particles completely. This results in a significant reduction of the Meissner effect when comparing with the bulk material, and it can be accounted by the following equation\cite{tinkham2004, Zolotavin2010}:
\begin{equation}
	\chi^{SC}=-\frac{3}{16}\frac{1}{40\pi}\frac{D^3}{\lambda_L^2 \xi_0}
\label{eq.1}
\end{equation}
where $D$ represents the particle diameter and $\xi_0\simeq 13$ nm is the coherence length for NbC\cite{Shang2020}. For particles with 8 nm diameter, eq. \ref{eq.1} leads to a diamagnetic susceptibility of $4\pi\chi^{SC}=37\times10^{-6}$ emu/cm$^3$Oe, consistent with the observations for the sample S4. However, a reduction of the Meissner susceptibility is noted upon decreasing the laser fluency (from S4 to S1 samples), which is attributed to an increase of defects other than C vacancies (i.e. NbC$_{1-x}$). On the other hand, $4\pi\chi^{SC}=0.7 \times10^{-3}$ emu/cm$^3$Oe for the P1 particles, which have an average size larger than $\lambda_L$. This value is also smaller than the expected value for bulk NbC, $4\pi\chi^{SC}=-1$. The observed $\chi^{SC}$ value for P1 suggests a particle size more akin to 22 nm, way smaller than the average particle size estimated ($\approx 470$ nm) and suggests the presence of numerous reminiscent defects in the crystalline structure, ultimately leading to a reduced superconducting volume. It is worth highlighting that the fabrication of nanoparticles through laser ablation in a liquid involves the detachment of hot atoms, vapors, and liquid drops followed by a rapid quenching of the material upon its interaction with the liquid\cite{Amendola2013}. This intricate process has the potential to generate a substantial amount of crystalline defects within the particle system. 

\begin{figure}
	\centering
	\includegraphics[width=\textwidth]{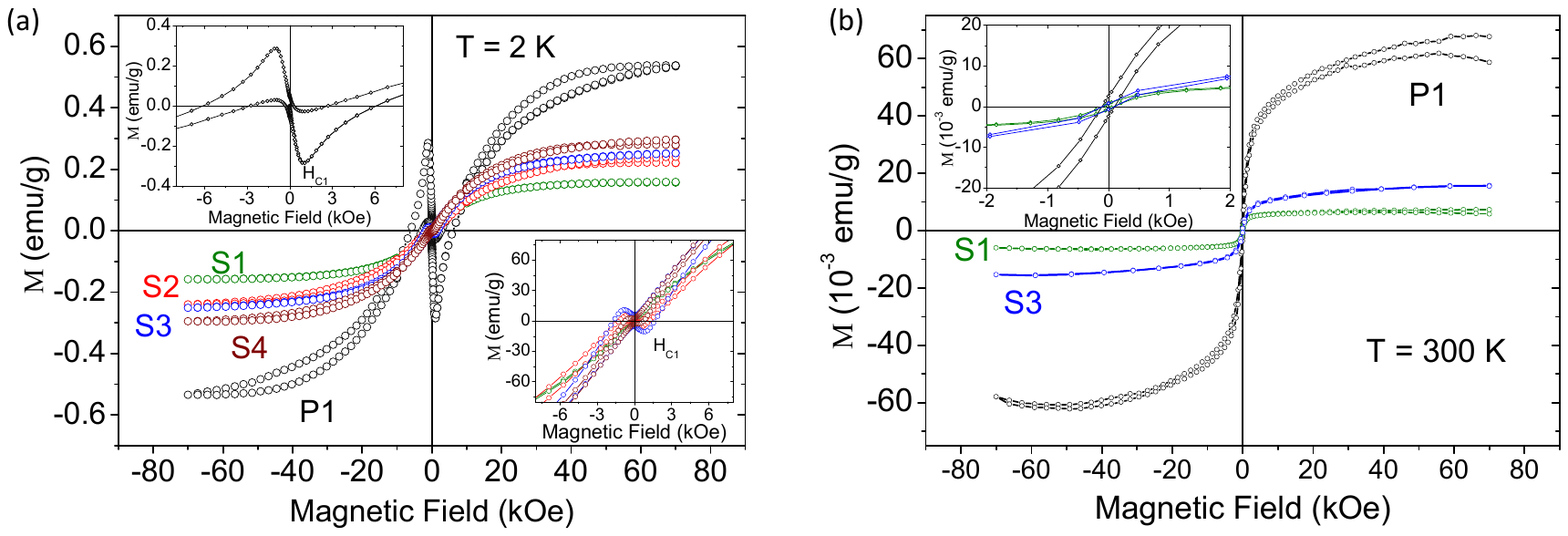}
	\caption{\label{Mag-MH}Magnetic hysteresis loops at 2K (a) and 300K (b). The insets in (a) provide a closer look at these curves at low field for the P1 sample (upper left) and S samples (lower right). Meanwhile, the inset in (b) is a magnified view at low field aimed at revealing the coercive field ($H_C$) of the order of 100 Oe at 300 K for this samples.}
\end{figure}

\begin{figure}[!htb]
	\centering
	\includegraphics[width=0.8\textwidth]{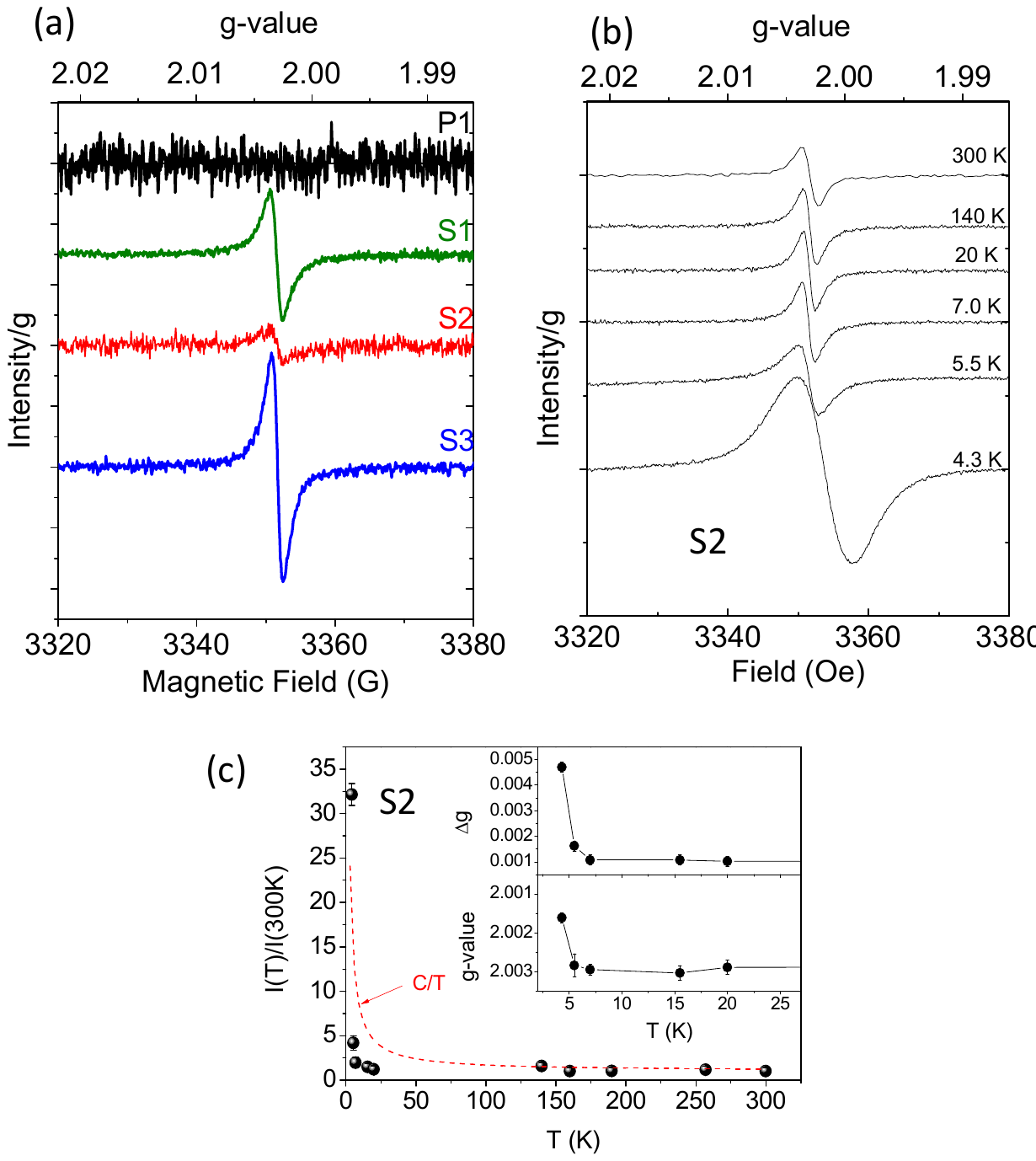}
	\caption{\label{ESR} (a) Electron spin resonance (ESR) spectra at 300 K for P1, S1, S2, and S3 samples.  In (b), ESR spectra for the S2 sample are presented at various temperatures. And in (c) the Normalized intensity $I(T)/I(300K)$ of the S2 samples as function of $T$. The red dashed line in (c) represent a Curie-like curve and inset the line width $\Delta H$ and resonance field $H_{res}$ at low $T$.}
\end{figure}

The isothermal magnetization as function of magnetic field at 2K and 300K for the NbC samples are shown in Figure \ref{Mag-MH}. The magnetization data was obtained after subtracting the diamagnetic contribution from the sample holder. Two distinct magnetic contribution are observed: a positive component and a superconducting loop at low temperature and low fields. The positive component, which is present even at room temperature, exhibits a larger magnetization for the P1 sample, decreasing with reduced fluency for the S samples. This positive magnetization with values on the order of $10^{-3}-10^{-2}$ emu/g and a coercitive field around $100$ Oe at room temperature has been reported for materials reduced to a nanometric scale\cite{SUNDARESAN2009,Li2008,Bose2005,Ma2022,Zhu2012,Jirsa2012,SHIPRA2013}. Surface ferromagnetism is suggested to be a universal feature in almost all nanoparticles of otherwise nonmagnetic inorganic materials, arising from surface cation or anion vacancies leading to ferromagnetic spin polarizations \cite{SUNDARESAN2009}. Moreover, since our P1 sample does not show any ESR signal (see the figure \ref{ESR}) we conclude that this ferromagnetic signal is ESR silent. Furthermore, the coexistence of superconductivity and ferromagnetism at low temperature has been observed in NPs of metal \cite{Li2008,Bose2005,Ma2022}, oxides\cite{Zhu2012,Jirsa2012} and nitrides\cite{SHIPRA2013}, but for the first time we report in carbides. However, it is acknowledged that this ferromagnetic signal may have originated from iron contamination in the synthesis process.

The electron spin resonance (ESR) experiments shown in Figure \ref{ESR}(a) present narrow resonances of line width $\Delta H\approx2$ G and a $g$-value at  $g=2.0$ only for the S samples. These resonances are often attributed to localized spins present in the samples and can not be attributed to the iron contamination. A quantification of the amount of spins against a standard sample indicated that there are only 1 spin every 10 to 20 nanoparticles for the S samples. However, within the accuracy of the ESR experiment the P1 sample does not present this ESR signal. This result indicates that this localized spin may be in the particles surface once the surface/volume rations for the S samples are about fifty times greater than that of the P1 sample. 

The $T$-dependence of the ESR spectra for the sample S2 is shown in the Figure \ref{ESR} (b). By lowering the temperature the signal do not change much, however below 10 K the signal starts to broaden, shift toward higher fields and its intensity increase. Figure \ref{ESR} (c) depicts the temperature dependence of the relative ESR intensity, $I(T)/I(300$K$)$. An illustrative Curie-like behavior reveals that the signal intensity displays an increase beyond this $T$-dependence. Futhermore, as highlighted in the inset of Figure \ref{ESR} (c), below $T_C$, the linewidth $\Delta H$ increases, and a small resonance shift toward higher fields is observed. Similar phenomena, including increased $\Delta H$ and $H_{res}$ below $T_C$, have been reported in bulk Type-II superconductors.\cite{Urbano2002} The line broadening is attributed to the field distribution within the superconducting vortex state, while the resonance shift is linked to the diamagnetism in the superconducting state. These results suggest that, in our samples, the localized spins on the particles' surfaces are being influenced by the diamagnetism of the superconducting vortex state in NbC particles, presumably through proximity superconducting effects.\cite{holm1932}

\section{Conclusions}

In conclusion, our study successfully demonstrated the synthesis of superconducting NbC nanoparticles through low-temperature laser ablation in acetone. Notably, varying the laser fluence showed negligible impact on the average size of the resulting NbC nanoparticles, which exhibited a broad size distribution. Some particles displayed surface features indicative of graphite or amorphous material, potentially porous amorphous carbon. The elevation in laser fluence at syntheses process increase the volume of the superconducting fraction. Additionally, a paramagnetic contribution was evident across all samples, even in the sedimented sample. Particularly noteworthy was the observation that this magnetic moment increased with rising fluence, especially notable in the supernatant sample at lower temperatures. These findings contribute valuable insights into the synthesis and properties of NbC nanoparticles, offering avenues for further exploration and application in superconductivity research.

\begin{acknowledgments}

This study was mainly supported and performed under the auspices of the S\~{a}o Paulo Research Foundation (FAPESP) through Grant No. 2017/10581-1 (Thematic). F.F. acknowledges his postdoc FAPESP Grant No. 2019/13678-1. R.R.U also acknowledges the National Council for Scientific and Technological Development (CNPq) for support through Grant No. 314587/2021-7. This research used facilities of the Brazilian Nanotechnology National Laboratory (LNNano), part of the Brazilian Centre for Research in Energy and Materials (CNPEM), a private non-profit organization under the supervision of the Brazilian Ministry for Science, Technology, and Innovations (MCTI). The electronic microscopy staff is acknowledged for the assistance during the experiments (proposal number: 20230481).

\end{acknowledgments}

\bibliography{ref}

\begin{thebibliography}{34}%
\makeatletter
\providecommand \@ifxundefined [1]{%
 \@ifx{#1\undefined}
}%
\providecommand \@ifnum [1]{%
 \ifnum #1\expandafter \@firstoftwo
 \else \expandafter \@secondoftwo
 \fi
}%
\providecommand \@ifx [1]{%
 \ifx #1\expandafter \@firstoftwo
 \else \expandafter \@secondoftwo
 \fi
}%
\providecommand \natexlab [1]{#1}%
\providecommand \enquote  [1]{``#1''}%
\providecommand \bibnamefont  [1]{#1}%
\providecommand \bibfnamefont [1]{#1}%
\providecommand \citenamefont [1]{#1}%
\providecommand \href@noop [0]{\@secondoftwo}%
\providecommand \href [0]{\begingroup \@sanitize@url \@href}%
\providecommand \@href[1]{\@@startlink{#1}\@@href}%
\providecommand \@@href[1]{\endgroup#1\@@endlink}%
\providecommand \@sanitize@url [0]{\catcode `\\12\catcode `\$12\catcode
  `\&12\catcode `\#12\catcode `\^12\catcode `\_12\catcode `\%12\relax}%
\providecommand \@@startlink[1]{}%
\providecommand \@@endlink[0]{}%
\providecommand \url  [0]{\begingroup\@sanitize@url \@url }%
\providecommand \@url [1]{\endgroup\@href {#1}{\urlprefix }}%
\providecommand \urlprefix  [0]{URL }%
\providecommand \Eprint [0]{\href }%
\providecommand \doibase [0]{http://dx.doi.org/}%
\providecommand \selectlanguage [0]{\@gobble}%
\providecommand \bibinfo  [0]{\@secondoftwo}%
\providecommand \bibfield  [0]{\@secondoftwo}%
\providecommand \translation [1]{[#1]}%
\providecommand \BibitemOpen [0]{}%
\providecommand \bibitemStop [0]{}%
\providecommand \bibitemNoStop [0]{.\EOS\space}%
\providecommand \EOS [0]{\spacefactor3000\relax}%
\providecommand \BibitemShut  [1]{\csname bibitem#1\endcsname}%
\let\auto@bib@innerbib\@empty
\bibitem [{\citenamefont {Cuppari}\ and\ \citenamefont
  {Santos}(2016)}]{Cuppari2016}%
  \BibitemOpen
  \bibfield  {author} {\bibinfo {author} {\bibfnamefont {M.~G. D.~V.}\
  \bibnamefont {Cuppari}}\ and\ \bibinfo {author} {\bibfnamefont {S.~F.}\
  \bibnamefont {Santos}},\ }\bibfield  {title} {\enquote {\bibinfo {title}
  {Physical properties of the nbc carbide},}\ }\href {\doibase
  10.3390/met6100250} {\bibfield  {journal} {\bibinfo  {journal} {Metals}\
  }\textbf {\bibinfo {volume} {6}} (\bibinfo {year} {2016}),\
  10.3390/met6100250}\BibitemShut {NoStop}%
\bibitem [{\citenamefont {Uhlmann}, \citenamefont {Meier},\ and\ \citenamefont
  {Hinzmann}(2019)}]{UHLMANN2019}%
  \BibitemOpen
  \bibfield  {author} {\bibinfo {author} {\bibfnamefont {E.}~\bibnamefont
  {Uhlmann}}, \bibinfo {author} {\bibfnamefont {P.}~\bibnamefont {Meier}}, \
  and\ \bibinfo {author} {\bibfnamefont {D.}~\bibnamefont {Hinzmann}},\
  }\bibfield  {title} {\enquote {\bibinfo {title} {Application of niobium
  carbide based cutting materials for peripheral milling of cfrp},}\ }\href
  {\doibase https://doi.org/10.1016/j.procir.2019.09.009} {\bibfield  {journal}
  {\bibinfo  {journal} {Procedia CIRP}\ }\textbf {\bibinfo {volume} {85}},\
  \bibinfo {pages} {108--113} (\bibinfo {year} {2019})}\BibitemShut {NoStop}%
\bibitem [{\citenamefont {Montenegro}\ \emph {et~al.}(2018)\citenamefont
  {Montenegro}, \citenamefont {Gomes}, \citenamefont {Rego},\ and\
  \citenamefont {Borille}}]{MONTENEGRO2018}%
  \BibitemOpen
  \bibfield  {author} {\bibinfo {author} {\bibfnamefont {P.}~\bibnamefont
  {Montenegro}}, \bibinfo {author} {\bibfnamefont {J.}~\bibnamefont {Gomes}},
  \bibinfo {author} {\bibfnamefont {R.}~\bibnamefont {Rego}}, \ and\ \bibinfo
  {author} {\bibfnamefont {A.}~\bibnamefont {Borille}},\ }\bibfield  {title}
  {\enquote {\bibinfo {title} {Potential of niobium carbide application as the
  hard phase in cutting tool substrate},}\ }\href {\doibase
  https://doi.org/10.1016/j.ijrmhm.2017.09.017} {\bibfield  {journal} {\bibinfo
   {journal} {International Journal of Refractory Metals and Hard Materials}\
  }\textbf {\bibinfo {volume} {70}},\ \bibinfo {pages} {116--123} (\bibinfo
  {year} {2018})}\BibitemShut {NoStop}%
\bibitem [{\citenamefont {Woydt}\ \emph {et~al.}(2016)\citenamefont {Woydt},
  \citenamefont {Mohrbacher}, \citenamefont {Vleugels},\ and\ \citenamefont
  {Huang}}]{WOYDT2016}%
  \BibitemOpen
  \bibfield  {author} {\bibinfo {author} {\bibfnamefont {M.}~\bibnamefont
  {Woydt}}, \bibinfo {author} {\bibfnamefont {H.}~\bibnamefont {Mohrbacher}},
  \bibinfo {author} {\bibfnamefont {J.}~\bibnamefont {Vleugels}}, \ and\
  \bibinfo {author} {\bibfnamefont {S.}~\bibnamefont {Huang}},\ }\bibfield
  {title} {\enquote {\bibinfo {title} {Niobium carbide for wear protection –
  tailoring its properties by processing and stoichiometry},}\ }\href {\doibase
  https://doi.org/10.1016/j.mprp.2015.12.010} {\bibfield  {journal} {\bibinfo
  {journal} {Metal Powder Report}\ }\textbf {\bibinfo {volume} {71}},\ \bibinfo
  {pages} {265--272} (\bibinfo {year} {2016})}\BibitemShut {NoStop}%
\bibitem [{\citenamefont {Woydt}\ and\ \citenamefont
  {Mohrbacher}(2013)}]{WOYDT2013}%
  \BibitemOpen
  \bibfield  {author} {\bibinfo {author} {\bibfnamefont {M.}~\bibnamefont
  {Woydt}}\ and\ \bibinfo {author} {\bibfnamefont {H.}~\bibnamefont
  {Mohrbacher}},\ }\bibfield  {title} {\enquote {\bibinfo {title} {Friction and
  wear of binder-less niobium carbide},}\ }\href {\doibase
  https://doi.org/10.1016/j.wear.2013.07.013} {\bibfield  {journal} {\bibinfo
  {journal} {Wear}\ }\textbf {\bibinfo {volume} {306}},\ \bibinfo {pages}
  {126--130} (\bibinfo {year} {2013})}\BibitemShut {NoStop}%
\bibitem [{\citenamefont {Alhowity}\ \emph {et~al.}(2024)\citenamefont
  {Alhowity}, \citenamefont {Balogun}, \citenamefont {Ganesan}, \citenamefont
  {Lund}, \citenamefont {Omolere}, \citenamefont {Adesope}, \citenamefont
  {Chukwunenye}, \citenamefont {Amagbor}, \citenamefont {Anwar}, \citenamefont
  {Altafi}, \citenamefont {D’Souza}, \citenamefont {Cundari},\ and\
  \citenamefont {Kelber}}]{Alhowity2024}%
  \BibitemOpen
  \bibfield  {author} {\bibinfo {author} {\bibfnamefont {S.}~\bibnamefont
  {Alhowity}}, \bibinfo {author} {\bibfnamefont {K.}~\bibnamefont {Balogun}},
  \bibinfo {author} {\bibfnamefont {A.}~\bibnamefont {Ganesan}}, \bibinfo
  {author} {\bibfnamefont {C.~J.}\ \bibnamefont {Lund}}, \bibinfo {author}
  {\bibfnamefont {O.}~\bibnamefont {Omolere}}, \bibinfo {author} {\bibfnamefont
  {Q.}~\bibnamefont {Adesope}}, \bibinfo {author} {\bibfnamefont
  {P.}~\bibnamefont {Chukwunenye}}, \bibinfo {author} {\bibfnamefont {S.~C.}\
  \bibnamefont {Amagbor}}, \bibinfo {author} {\bibfnamefont {F.}~\bibnamefont
  {Anwar}}, \bibinfo {author} {\bibfnamefont {M.~K.}\ \bibnamefont {Altafi}},
  \bibinfo {author} {\bibfnamefont {F.}~\bibnamefont {D’Souza}}, \bibinfo
  {author} {\bibfnamefont {T.~R.}\ \bibnamefont {Cundari}}, \ and\ \bibinfo
  {author} {\bibfnamefont {J.~A.}\ \bibnamefont {Kelber}},\ }\bibfield  {title}
  {\enquote {\bibinfo {title} {Niobium carbide and tantalum carbide as nitrogen
  reduction electrocatalysts: Catalytic activity, carbophilicity, and the
  importance of intermediate oxidation states},}\ }\href {\doibase
  10.1021/acsami.3c11683} {\bibfield  {journal} {\bibinfo  {journal} {ACS
  Applied Materials \& Interfaces}\ }\textbf {\bibinfo {volume} {16}},\
  \bibinfo {pages} {2180--2192} (\bibinfo {year} {2024})}\BibitemShut {NoStop}%
\bibitem [{\citenamefont {Yang}\ \emph {et~al.}(2021)\citenamefont {Yang},
  \citenamefont {Zhao}, \citenamefont {Liao}, \citenamefont {Cheng},
  \citenamefont {Mao}, \citenamefont {Fa},\ and\ \citenamefont
  {Chen}}]{Yang2021}%
  \BibitemOpen
  \bibfield  {author} {\bibinfo {author} {\bibfnamefont {G.}~\bibnamefont
  {Yang}}, \bibinfo {author} {\bibfnamefont {X.}~\bibnamefont {Zhao}}, \bibinfo
  {author} {\bibfnamefont {F.}~\bibnamefont {Liao}}, \bibinfo {author}
  {\bibfnamefont {Q.}~\bibnamefont {Cheng}}, \bibinfo {author} {\bibfnamefont
  {L.}~\bibnamefont {Mao}}, \bibinfo {author} {\bibfnamefont {H.}~\bibnamefont
  {Fa}}, \ and\ \bibinfo {author} {\bibfnamefont {L.}~\bibnamefont {Chen}},\
  }\bibfield  {title} {\enquote {\bibinfo {title} {Recent progress and
  applications of niobium-based nanomaterials and their composites for
  supercapacitors and hybrid ion capacitors},}\ }\href {\doibase
  10.1039/D1SE00397F} {\bibfield  {journal} {\bibinfo  {journal} {Sustainable
  Energy Fuels}\ }\textbf {\bibinfo {volume} {5}},\ \bibinfo {pages}
  {3039--3083} (\bibinfo {year} {2021})}\BibitemShut {NoStop}%
\bibitem [{\citenamefont {Bhattacharjee}, \citenamefont {Pati},\ and\
  \citenamefont {Maity}(2016)}]{Kaustav2016}%
  \BibitemOpen
  \bibfield  {author} {\bibinfo {author} {\bibfnamefont {K.}~\bibnamefont
  {Bhattacharjee}}, \bibinfo {author} {\bibfnamefont {S.~P.}\ \bibnamefont
  {Pati}}, \ and\ \bibinfo {author} {\bibfnamefont {A.}~\bibnamefont {Maity}},\
  }\bibfield  {title} {\enquote {\bibinfo {title} {High critical field nbc
  superconductor on carbon spheres},}\ }\href {\doibase 10.1039/C6CP01771A}
  {\bibfield  {journal} {\bibinfo  {journal} {Phys. Chem. Chem. Phys.}\
  }\textbf {\bibinfo {volume} {18}},\ \bibinfo {pages} {15218--15222} (\bibinfo
  {year} {2016})}\BibitemShut {NoStop}%
\bibitem [{\citenamefont {Zou}\ \emph {et~al.}(2011)\citenamefont {Zou},
  \citenamefont {Luo}, \citenamefont {Baily}, \citenamefont {Zhang},
  \citenamefont {Haberkorn}, \citenamefont {Xiong}, \citenamefont {Bauer},
  \citenamefont {McCleskey}, \citenamefont {Burrell}, \citenamefont {Civale}
  \emph {et~al.}}]{zou2011}%
  \BibitemOpen
  \bibfield  {author} {\bibinfo {author} {\bibfnamefont {G.}~\bibnamefont
  {Zou}}, \bibinfo {author} {\bibfnamefont {H.}~\bibnamefont {Luo}}, \bibinfo
  {author} {\bibfnamefont {S.}~\bibnamefont {Baily}}, \bibinfo {author}
  {\bibfnamefont {Y.}~\bibnamefont {Zhang}}, \bibinfo {author} {\bibfnamefont
  {N.}~\bibnamefont {Haberkorn}}, \bibinfo {author} {\bibfnamefont
  {J.}~\bibnamefont {Xiong}}, \bibinfo {author} {\bibfnamefont
  {E.}~\bibnamefont {Bauer}}, \bibinfo {author} {\bibfnamefont
  {T.}~\bibnamefont {McCleskey}}, \bibinfo {author} {\bibfnamefont
  {A.}~\bibnamefont {Burrell}}, \bibinfo {author} {\bibfnamefont
  {L.}~\bibnamefont {Civale}},  \emph {et~al.},\ }\bibfield  {title} {\enquote
  {\bibinfo {title} {Highly aligned carbon nanotube forests coated by
  superconducting nbc},}\ }\href {\doibase 10.1038/ncomms1438} {\bibfield
  {journal} {\bibinfo  {journal} {Nature Communications}\ }\textbf {\bibinfo
  {volume} {2}},\ \bibinfo {pages} {428} (\bibinfo {year} {2011})}\BibitemShut
  {NoStop}%
\bibitem [{\citenamefont {Zhang}\ \emph {et~al.}(2012)\citenamefont {Zhang},
  \citenamefont {Ronning}, \citenamefont {Gofryk}, \citenamefont {Mara},
  \citenamefont {Haberkorn}, \citenamefont {Zou}, \citenamefont {Wang},
  \citenamefont {Lee}, \citenamefont {Bauer}, \citenamefont {McCleskey},
  \citenamefont {Burell}, \citenamefont {Civale}, \citenamefont {Zhu},\ and\
  \citenamefont {Jia}}]{Yingying2012}%
  \BibitemOpen
  \bibfield  {author} {\bibinfo {author} {\bibfnamefont {Y.}~\bibnamefont
  {Zhang}}, \bibinfo {author} {\bibfnamefont {F.}~\bibnamefont {Ronning}},
  \bibinfo {author} {\bibfnamefont {K.}~\bibnamefont {Gofryk}}, \bibinfo
  {author} {\bibfnamefont {N.~A.}\ \bibnamefont {Mara}}, \bibinfo {author}
  {\bibfnamefont {N.}~\bibnamefont {Haberkorn}}, \bibinfo {author}
  {\bibfnamefont {G.}~\bibnamefont {Zou}}, \bibinfo {author} {\bibfnamefont
  {H.}~\bibnamefont {Wang}}, \bibinfo {author} {\bibfnamefont {J.~H.}\
  \bibnamefont {Lee}}, \bibinfo {author} {\bibfnamefont {E.}~\bibnamefont
  {Bauer}}, \bibinfo {author} {\bibfnamefont {T.~M.}\ \bibnamefont
  {McCleskey}}, \bibinfo {author} {\bibfnamefont {A.~K.}\ \bibnamefont
  {Burell}}, \bibinfo {author} {\bibfnamefont {L.}~\bibnamefont {Civale}},
  \bibinfo {author} {\bibfnamefont {Y.~T.}\ \bibnamefont {Zhu}}, \ and\
  \bibinfo {author} {\bibfnamefont {Q.}~\bibnamefont {Jia}},\ }\bibfield
  {title} {\enquote {\bibinfo {title} {Aligned carbon nanotubes sandwiched in
  epitaxial nbc film for enhanced superconductivity},}\ }\href {\doibase
  10.1039/C2NR11906D} {\bibfield  {journal} {\bibinfo  {journal} {Nanoscale}\
  }\textbf {\bibinfo {volume} {4}},\ \bibinfo {pages} {2268--2271} (\bibinfo
  {year} {2012})}\BibitemShut {NoStop}%
\bibitem [{\citenamefont {Wells}\ \emph {et~al.}(1964)\citenamefont {Wells},
  \citenamefont {Pickus}, \citenamefont {Kennedy},\ and\ \citenamefont
  {Zackay}}]{Wells1964}%
  \BibitemOpen
  \bibfield  {author} {\bibinfo {author} {\bibfnamefont {M.}~\bibnamefont
  {Wells}}, \bibinfo {author} {\bibfnamefont {M.}~\bibnamefont {Pickus}},
  \bibinfo {author} {\bibfnamefont {K.}~\bibnamefont {Kennedy}}, \ and\
  \bibinfo {author} {\bibfnamefont {V.}~\bibnamefont {Zackay}},\ }\bibfield
  {title} {\enquote {\bibinfo {title} {Superconductivity of solid solutions of
  tac and nbc},}\ }\href {\doibase 10.1103/PhysRevLett.12.536} {\bibfield
  {journal} {\bibinfo  {journal} {Phys. Rev. Lett.}\ }\textbf {\bibinfo
  {volume} {12}},\ \bibinfo {pages} {536--538} (\bibinfo {year}
  {1964})}\BibitemShut {NoStop}%
\bibitem [{\citenamefont {Gusev}\ and\ \citenamefont
  {Rempel}(1989)}]{Gusev1989}%
  \BibitemOpen
  \bibfield  {author} {\bibinfo {author} {\bibfnamefont {A.~I.}\ \bibnamefont
  {Gusev}}\ and\ \bibinfo {author} {\bibfnamefont {A.~A.}\ \bibnamefont
  {Rempel}},\ }\bibfield  {title} {\enquote {\bibinfo {title}
  {Superconductivity in disordered and ordered niobium carbide},}\ }\href
  {\doibase https://doi.org/10.1002/pssb.2221510125} {\bibfield  {journal}
  {\bibinfo  {journal} {physica status solidi (b)}\ }\textbf {\bibinfo {volume}
  {151}},\ \bibinfo {pages} {211--224} (\bibinfo {year} {1989})}\BibitemShut
  {NoStop}%
\bibitem [{\citenamefont {Li}\ \emph {et~al.}(2006)\citenamefont {Li},
  \citenamefont {Li}, \citenamefont {Ma},\ and\ \citenamefont
  {Zhang}}]{Li2006}%
  \BibitemOpen
  \bibfield  {author} {\bibinfo {author} {\bibfnamefont {D.}~\bibnamefont
  {Li}}, \bibinfo {author} {\bibfnamefont {W.~F.}\ \bibnamefont {Li}}, \bibinfo
  {author} {\bibfnamefont {S.}~\bibnamefont {Ma}}, \ and\ \bibinfo {author}
  {\bibfnamefont {Z.~D.}\ \bibnamefont {Zhang}},\ }\bibfield  {title} {\enquote
  {\bibinfo {title} {Electronic transport properties of nbc(c)-c
  nanocomposites},}\ }\href {\doibase 10.1103/PhysRevB.73.193402} {\bibfield
  {journal} {\bibinfo  {journal} {Phys. Rev. B}\ }\textbf {\bibinfo {volume}
  {73}},\ \bibinfo {pages} {193402} (\bibinfo {year} {2006})}\BibitemShut
  {NoStop}%
\bibitem [{\citenamefont {Fukunaga}, \citenamefont {Chu},\ and\ \citenamefont
  {McHenry}(1998)}]{Fukunaga1998}%
  \BibitemOpen
  \bibfield  {author} {\bibinfo {author} {\bibfnamefont {A.}~\bibnamefont
  {Fukunaga}}, \bibinfo {author} {\bibfnamefont {S.}~\bibnamefont {Chu}}, \
  and\ \bibinfo {author} {\bibfnamefont {M.~E.}\ \bibnamefont {McHenry}},\
  }\bibfield  {title} {\enquote {\bibinfo {title} {Synthesis, structure, and
  superconducting properties of tantalum carbide nanorods and nanoparticles},}\
  }\href {\doibase 10.1557/JMR.1998.0345} {\bibfield  {journal} {\bibinfo
  {journal} {Journal of Materials Research}\ }\textbf {\bibinfo {volume}
  {13}},\ \bibinfo {pages} {2465–2471} (\bibinfo {year} {1998})}\BibitemShut
  {NoStop}%
\bibitem [{\citenamefont {Zhang}\ \emph {et~al.}(2019)\citenamefont {Zhang},
  \citenamefont {Zhang}, \citenamefont {Liu}, \citenamefont {Chen},
  \citenamefont {Zhu},\ and\ \citenamefont {Liang}}]{Zhang2019}%
  \BibitemOpen
  \bibfield  {author} {\bibinfo {author} {\bibfnamefont {D.}~\bibnamefont
  {Zhang}}, \bibinfo {author} {\bibfnamefont {C.}~\bibnamefont {Zhang}},
  \bibinfo {author} {\bibfnamefont {J.}~\bibnamefont {Liu}}, \bibinfo {author}
  {\bibfnamefont {Q.}~\bibnamefont {Chen}}, \bibinfo {author} {\bibfnamefont
  {X.}~\bibnamefont {Zhu}}, \ and\ \bibinfo {author} {\bibfnamefont
  {C.}~\bibnamefont {Liang}},\ }\bibfield  {title} {\enquote {\bibinfo {title}
  {Carbon-encapsulated metal/metal carbide/metal oxide core–shell
  nanostructures generated by laser ablation of metals in organic solvents},}\
  }\href {\doibase 10.1021/acsanm.8b01541} {\bibfield  {journal} {\bibinfo
  {journal} {ACS Applied Nano Materials}\ }\textbf {\bibinfo {volume} {2}},\
  \bibinfo {pages} {28--39} (\bibinfo {year} {2019})}\BibitemShut {NoStop}%
\bibitem [{\citenamefont {Zhang}\ \emph {et~al.}(2016)\citenamefont {Zhang},
  \citenamefont {Liu}, \citenamefont {Tian}, \citenamefont {Ye}, \citenamefont
  {Cai}, \citenamefont {Liang},\ and\ \citenamefont {Terabe}}]{ZHANG2016}%
  \BibitemOpen
  \bibfield  {author} {\bibinfo {author} {\bibfnamefont {H.}~\bibnamefont
  {Zhang}}, \bibinfo {author} {\bibfnamefont {J.}~\bibnamefont {Liu}}, \bibinfo
  {author} {\bibfnamefont {Z.}~\bibnamefont {Tian}}, \bibinfo {author}
  {\bibfnamefont {Y.}~\bibnamefont {Ye}}, \bibinfo {author} {\bibfnamefont
  {Y.}~\bibnamefont {Cai}}, \bibinfo {author} {\bibfnamefont {C.}~\bibnamefont
  {Liang}}, \ and\ \bibinfo {author} {\bibfnamefont {K.}~\bibnamefont
  {Terabe}},\ }\bibfield  {title} {\enquote {\bibinfo {title} {A general
  strategy toward transition metal carbide/carbon core/shell nanospheres and
  their application for supercapacitor electrode},}\ }\href {\doibase
  https://doi.org/10.1016/j.carbon.2016.01.047} {\bibfield  {journal} {\bibinfo
   {journal} {Carbon}\ }\textbf {\bibinfo {volume} {100}},\ \bibinfo {pages}
  {590--599} (\bibinfo {year} {2016})}\BibitemShut {NoStop}%
\bibitem [{\citenamefont {Amendola}\ and\ \citenamefont
  {Meneghetti}(2013)}]{Amendola2013}%
  \BibitemOpen
  \bibfield  {author} {\bibinfo {author} {\bibfnamefont {V.}~\bibnamefont
  {Amendola}}\ and\ \bibinfo {author} {\bibfnamefont {M.}~\bibnamefont
  {Meneghetti}},\ }\bibfield  {title} {\enquote {\bibinfo {title} {What
  controls the composition and the structure of nanomaterials generated by
  laser ablation in liquid solution?}}\ }\href {\doibase 10.1039/C2CP42895D}
  {\bibfield  {journal} {\bibinfo  {journal} {Phys. Chem. Chem. Phys.}\
  }\textbf {\bibinfo {volume} {15}},\ \bibinfo {pages} {3027--3046} (\bibinfo
  {year} {2013})}\BibitemShut {NoStop}%
\bibitem [{\citenamefont {Frias~Batista}\ \emph {et~al.}(2022)\citenamefont
  {Frias~Batista}, \citenamefont {Nag}, \citenamefont {Meader},\ and\
  \citenamefont {Tibbetts}}]{Frias2022}%
  \BibitemOpen
  \bibfield  {author} {\bibinfo {author} {\bibfnamefont {L.~M.}\ \bibnamefont
  {Frias~Batista}}, \bibinfo {author} {\bibfnamefont {A.}~\bibnamefont {Nag}},
  \bibinfo {author} {\bibfnamefont {V.~K.}\ \bibnamefont {Meader}}, \ and\
  \bibinfo {author} {\bibfnamefont {K.~M.}\ \bibnamefont {Tibbetts}},\
  }\bibfield  {title} {\enquote {\bibinfo {title} {Generation of nanomaterials
  by reactive laser-synthesis in liquid},}\ }\href {\doibase
  10.1007/s11433-021-1835-x} {\bibfield  {journal} {\bibinfo  {journal}
  {Science China: Physics, Mechanics and Astronomy}\ }\textbf {\bibinfo
  {volume} {65}} (\bibinfo {year} {2022}),\
  10.1007/s11433-021-1835-x}\BibitemShut {NoStop}%
\bibitem [{\citenamefont {Zhang}\ \emph {et~al.}(2013)\citenamefont {Zhang},
  \citenamefont {Liang}, \citenamefont {Liu}, \citenamefont {Tian},\ and\
  \citenamefont {Shao}}]{ZHANG2013}%
  \BibitemOpen
  \bibfield  {author} {\bibinfo {author} {\bibfnamefont {H.}~\bibnamefont
  {Zhang}}, \bibinfo {author} {\bibfnamefont {C.}~\bibnamefont {Liang}},
  \bibinfo {author} {\bibfnamefont {J.}~\bibnamefont {Liu}}, \bibinfo {author}
  {\bibfnamefont {Z.}~\bibnamefont {Tian}}, \ and\ \bibinfo {author}
  {\bibfnamefont {G.}~\bibnamefont {Shao}},\ }\bibfield  {title} {\enquote
  {\bibinfo {title} {The formation of onion-like carbon-encapsulated cobalt
  carbide core/shell nanoparticles by the laser ablation of metallic cobalt in
  acetone},}\ }\href {\doibase https://doi.org/10.1016/j.carbon.2012.12.015}
  {\bibfield  {journal} {\bibinfo  {journal} {Carbon}\ }\textbf {\bibinfo
  {volume} {55}},\ \bibinfo {pages} {108--115} (\bibinfo {year}
  {2013})}\BibitemShut {NoStop}%
\bibitem [{\citenamefont {A}\ \emph {et~al.}(2020)\citenamefont {A},
  \citenamefont {M}, \citenamefont {A}, \citenamefont {A},\ and\ \citenamefont
  {R.}}]{Debonis2020}%
  \BibitemOpen
  \bibfield  {author} {\bibinfo {author} {\bibfnamefont {D.~B.}\ \bibnamefont
  {A}}, \bibinfo {author} {\bibfnamefont {C.}~\bibnamefont {M}}, \bibinfo
  {author} {\bibfnamefont {S.}~\bibnamefont {A}}, \bibinfo {author}
  {\bibfnamefont {G.}~\bibnamefont {A}}, \ and\ \bibinfo {author}
  {\bibfnamefont {T.}~\bibnamefont {R.}},\ }\bibfield  {title} {\enquote
  {\bibinfo {title} {Transition metal carbide core/shell nanoparticles by
  ultra-short laser ablation in liquid.}}\ }\href {\doibase
  10.3390/nano10010145} {\bibfield  {journal} {\bibinfo  {journal}
  {Nanomaterials}\ }\textbf {\bibinfo {volume} {14}},\ \bibinfo {pages} {145)}
  (\bibinfo {year} {2020})}\BibitemShut {NoStop}%
\bibitem [{\citenamefont {Nemanich}\ and\ \citenamefont
  {Solin}(1979)}]{Nemanich1979}%
  \BibitemOpen
  \bibfield  {author} {\bibinfo {author} {\bibfnamefont {R.~J.}\ \bibnamefont
  {Nemanich}}\ and\ \bibinfo {author} {\bibfnamefont {S.~A.}\ \bibnamefont
  {Solin}},\ }\bibfield  {title} {\enquote {\bibinfo {title} {First- and
  second-order raman scattering from finite-size crystals of graphite},}\
  }\href {\doibase 10.1103/PhysRevB.20.392} {\bibfield  {journal} {\bibinfo
  {journal} {Phys. Rev. B}\ }\textbf {\bibinfo {volume} {20}},\ \bibinfo
  {pages} {392--401} (\bibinfo {year} {1979})}\BibitemShut {NoStop}%
\bibitem [{\citenamefont {Giorgi}\ \emph {et~al.}(1962)\citenamefont {Giorgi},
  \citenamefont {Szklarz}, \citenamefont {Storms}, \citenamefont {Bowman},\
  and\ \citenamefont {Matthias}}]{Giorgi1962}%
  \BibitemOpen
  \bibfield  {author} {\bibinfo {author} {\bibfnamefont {A.~L.}\ \bibnamefont
  {Giorgi}}, \bibinfo {author} {\bibfnamefont {E.~G.}\ \bibnamefont {Szklarz}},
  \bibinfo {author} {\bibfnamefont {E.~K.}\ \bibnamefont {Storms}}, \bibinfo
  {author} {\bibfnamefont {A.~L.}\ \bibnamefont {Bowman}}, \ and\ \bibinfo
  {author} {\bibfnamefont {B.~T.}\ \bibnamefont {Matthias}},\ }\bibfield
  {title} {\enquote {\bibinfo {title} {Effect of composition on the
  superconducting transition temperature of tantalum carbide and niobium
  carbide},}\ }\href {\doibase 10.1103/PhysRev.125.837} {\bibfield  {journal}
  {\bibinfo  {journal} {Phys. Rev.}\ }\textbf {\bibinfo {volume} {125}},\
  \bibinfo {pages} {837--838} (\bibinfo {year} {1962})}\BibitemShut {NoStop}%
\bibitem [{\citenamefont {Shang}\ \emph {et~al.}(2020)\citenamefont {Shang},
  \citenamefont {Zhao}, \citenamefont {Gawryluk}, \citenamefont {Shi},
  \citenamefont {Medarde}, \citenamefont {Pomjakushina},\ and\ \citenamefont
  {Shiroka}}]{Shang2020}%
  \BibitemOpen
  \bibfield  {author} {\bibinfo {author} {\bibfnamefont {T.}~\bibnamefont
  {Shang}}, \bibinfo {author} {\bibfnamefont {J.~Z.}\ \bibnamefont {Zhao}},
  \bibinfo {author} {\bibfnamefont {D.~J.}\ \bibnamefont {Gawryluk}}, \bibinfo
  {author} {\bibfnamefont {M.}~\bibnamefont {Shi}}, \bibinfo {author}
  {\bibfnamefont {M.}~\bibnamefont {Medarde}}, \bibinfo {author} {\bibfnamefont
  {E.}~\bibnamefont {Pomjakushina}}, \ and\ \bibinfo {author} {\bibfnamefont
  {T.}~\bibnamefont {Shiroka}},\ }\bibfield  {title} {\enquote {\bibinfo
  {title} {Superconductivity and topological aspects of the rocksalt carbides
  nbc and tac},}\ }\href {\doibase 10.1103/PhysRevB.101.214518} {\bibfield
  {journal} {\bibinfo  {journal} {Phys. Rev. B}\ }\textbf {\bibinfo {volume}
  {101}},\ \bibinfo {pages} {214518} (\bibinfo {year} {2020})}\BibitemShut
  {NoStop}%
\bibitem [{\citenamefont {Tinkham}(2004)}]{tinkham2004}%
  \BibitemOpen
  \bibfield  {author} {\bibinfo {author} {\bibfnamefont {M.}~\bibnamefont
  {Tinkham}},\ }\href@noop {} {\emph {\bibinfo {title} {Introduction to
  superconductivity}}}\ (\bibinfo  {publisher} {Courier Corporation},\ \bibinfo
  {year} {2004})\BibitemShut {NoStop}%
\bibitem [{\citenamefont {Zolotavin}\ and\ \citenamefont
  {Guyot-Sionnest}(2010)}]{Zolotavin2010}%
  \BibitemOpen
  \bibfield  {author} {\bibinfo {author} {\bibfnamefont {P.}~\bibnamefont
  {Zolotavin}}\ and\ \bibinfo {author} {\bibfnamefont {P.}~\bibnamefont
  {Guyot-Sionnest}},\ }\bibfield  {title} {\enquote {\bibinfo {title} {Meissner
  effect in colloidal pb nanoparticles},}\ }\href {\doibase 10.1021/nn102009g}
  {\bibfield  {journal} {\bibinfo  {journal} {ACS Nano}\ }\textbf {\bibinfo
  {volume} {4}},\ \bibinfo {pages} {5599--5608} (\bibinfo {year}
  {2010})}\BibitemShut {NoStop}%
\bibitem [{\citenamefont {Sundaresan}\ and\ \citenamefont
  {Rao}(2009)}]{SUNDARESAN2009}%
  \BibitemOpen
  \bibfield  {author} {\bibinfo {author} {\bibfnamefont {A.}~\bibnamefont
  {Sundaresan}}\ and\ \bibinfo {author} {\bibfnamefont {C.}~\bibnamefont
  {Rao}},\ }\bibfield  {title} {\enquote {\bibinfo {title} {Ferromagnetism as a
  universal feature of inorganic nanoparticles},}\ }\href {\doibase
  https://doi.org/10.1016/j.nantod.2008.10.002} {\bibfield  {journal} {\bibinfo
   {journal} {Nano Today}\ }\textbf {\bibinfo {volume} {4}},\ \bibinfo {pages}
  {96--106} (\bibinfo {year} {2009})}\BibitemShut {NoStop}%
\bibitem [{\citenamefont {Li}\ \emph {et~al.}(2008)\citenamefont {Li},
  \citenamefont {Wang}, \citenamefont {Li}, \citenamefont {Hsu}, \citenamefont
  {Yang},\ and\ \citenamefont {Wu}}]{Li2008}%
  \BibitemOpen
  \bibfield  {author} {\bibinfo {author} {\bibfnamefont {W.-H.}\ \bibnamefont
  {Li}}, \bibinfo {author} {\bibfnamefont {C.-W.}\ \bibnamefont {Wang}},
  \bibinfo {author} {\bibfnamefont {C.-Y.}\ \bibnamefont {Li}}, \bibinfo
  {author} {\bibfnamefont {C.~K.}\ \bibnamefont {Hsu}}, \bibinfo {author}
  {\bibfnamefont {C.~C.}\ \bibnamefont {Yang}}, \ and\ \bibinfo {author}
  {\bibfnamefont {C.-M.}\ \bibnamefont {Wu}},\ }\bibfield  {title} {\enquote
  {\bibinfo {title} {Coexistence of ferromagnetism and superconductivity in sn
  nanoparticles},}\ }\href {\doibase 10.1103/PhysRevB.77.094508} {\bibfield
  {journal} {\bibinfo  {journal} {Phys. Rev. B}\ }\textbf {\bibinfo {volume}
  {77}},\ \bibinfo {pages} {094508} (\bibinfo {year} {2008})}\BibitemShut
  {NoStop}%
\bibitem [{\citenamefont {Bose}\ \emph {et~al.}(2005)\citenamefont {Bose},
  \citenamefont {Raychaudhuri}, \citenamefont {Banerjee}, \citenamefont
  {Vasa},\ and\ \citenamefont {Ayyub}}]{Bose2005}%
  \BibitemOpen
  \bibfield  {author} {\bibinfo {author} {\bibfnamefont {S.}~\bibnamefont
  {Bose}}, \bibinfo {author} {\bibfnamefont {P.}~\bibnamefont {Raychaudhuri}},
  \bibinfo {author} {\bibfnamefont {R.}~\bibnamefont {Banerjee}}, \bibinfo
  {author} {\bibfnamefont {P.}~\bibnamefont {Vasa}}, \ and\ \bibinfo {author}
  {\bibfnamefont {P.}~\bibnamefont {Ayyub}},\ }\bibfield  {title} {\enquote
  {\bibinfo {title} {Mechanism of the size dependence of the superconducting
  transition of nanostructured nb},}\ }\href {\doibase
  10.1103/PhysRevLett.95.147003} {\bibfield  {journal} {\bibinfo  {journal}
  {Phys. Rev. Lett.}\ }\textbf {\bibinfo {volume} {95}},\ \bibinfo {pages}
  {147003} (\bibinfo {year} {2005})}\BibitemShut {NoStop}%
\bibitem [{\citenamefont {Ma}\ \emph {et~al.}(2022)\citenamefont {Ma},
  \citenamefont {Batsaikhan}, \citenamefont {Chen}, \citenamefont {Chen},
  \citenamefont {Lee}, \citenamefont {Li}, \citenamefont {Wu},\ and\
  \citenamefont {Wang}}]{Ma2022}%
  \BibitemOpen
  \bibfield  {author} {\bibinfo {author} {\bibfnamefont {M.-H.}\ \bibnamefont
  {Ma}}, \bibinfo {author} {\bibfnamefont {E.}~\bibnamefont {Batsaikhan}},
  \bibinfo {author} {\bibfnamefont {H.-N.}\ \bibnamefont {Chen}}, \bibinfo
  {author} {\bibfnamefont {T.-Y.}\ \bibnamefont {Chen}}, \bibinfo {author}
  {\bibfnamefont {C.-H.}\ \bibnamefont {Lee}}, \bibinfo {author} {\bibfnamefont
  {W.-H.}\ \bibnamefont {Li}}, \bibinfo {author} {\bibfnamefont {C.-M.}\
  \bibnamefont {Wu}}, \ and\ \bibinfo {author} {\bibfnamefont {C.-W.}\
  \bibnamefont {Wang}},\ }\bibfield  {title} {\enquote {\bibinfo {title}
  {Non-conventional superconductivity in magnetic in and sn nanoparticles},}\
  }\href {\doibase 10.1038/s41598-022-04889-6} {\bibfield  {journal} {\bibinfo
  {journal} {Scientific Reports}\ }\textbf {\bibinfo {volume} {12}},\ \bibinfo
  {pages} {775} (\bibinfo {year} {2022})}\BibitemShut {NoStop}%
\bibitem [{\citenamefont {Zhu}\ \emph {et~al.}(2012)\citenamefont {Zhu},
  \citenamefont {Gao}, \citenamefont {Dong}, \citenamefont {Yang},
  \citenamefont {Zhang}, \citenamefont {Zhang}, \citenamefont {Shi},
  \citenamefont {Gao}, \citenamefont {Luo},\ and\ \citenamefont
  {Xue}}]{Zhu2012}%
  \BibitemOpen
  \bibfield  {author} {\bibinfo {author} {\bibfnamefont {Z.}~\bibnamefont
  {Zhu}}, \bibinfo {author} {\bibfnamefont {D.}~\bibnamefont {Gao}}, \bibinfo
  {author} {\bibfnamefont {C.}~\bibnamefont {Dong}}, \bibinfo {author}
  {\bibfnamefont {G.}~\bibnamefont {Yang}}, \bibinfo {author} {\bibfnamefont
  {J.}~\bibnamefont {Zhang}}, \bibinfo {author} {\bibfnamefont
  {J.}~\bibnamefont {Zhang}}, \bibinfo {author} {\bibfnamefont
  {Z.}~\bibnamefont {Shi}}, \bibinfo {author} {\bibfnamefont {H.}~\bibnamefont
  {Gao}}, \bibinfo {author} {\bibfnamefont {H.}~\bibnamefont {Luo}}, \ and\
  \bibinfo {author} {\bibfnamefont {D.}~\bibnamefont {Xue}},\ }\bibfield
  {title} {\enquote {\bibinfo {title} {Coexistence of ferromagnetism and
  superconductivity in ybco nanoparticles},}\ }\href {\doibase
  10.1039/C2CP23046A} {\bibfield  {journal} {\bibinfo  {journal} {Phys. Chem.
  Chem. Phys.}\ }\textbf {\bibinfo {volume} {14}},\ \bibinfo {pages}
  {3859--3863} (\bibinfo {year} {2012})}\BibitemShut {NoStop}%
\bibitem [{\citenamefont {Jirsa}, \citenamefont {Rames},\ and\ \citenamefont
  {Wolf}(2012)}]{Jirsa2012}%
  \BibitemOpen
  \bibfield  {author} {\bibinfo {author} {\bibfnamefont {M.}~\bibnamefont
  {Jirsa}}, \bibinfo {author} {\bibfnamefont {M.}~\bibnamefont {Rames}}, \ and\
  \bibinfo {author} {\bibfnamefont {T.}~\bibnamefont {Wolf}},\ }\bibfield
  {title} {\enquote {\bibinfo {title} {Interplay of paramagnetic signal with
  superconductive environment in a (nd,eu,gd)bacuo single crystal},}\ }\href
  {\doibase 10.1088/1742-6596/400/2/022039} {\bibfield  {journal} {\bibinfo
  {journal} {Journal of Physics: Conference Series}\ }\textbf {\bibinfo
  {volume} {400}},\ \bibinfo {pages} {022039} (\bibinfo {year}
  {2012})}\BibitemShut {NoStop}%
\bibitem [{\citenamefont {Shipra}, \citenamefont {Kumar},\ and\ \citenamefont
  {Sundaresan}(2013)}]{SHIPRA2013}%
  \BibitemOpen
  \bibfield  {author} {\bibinfo {author} {\bibfnamefont {R.}~\bibnamefont
  {Shipra}}, \bibinfo {author} {\bibfnamefont {N.}~\bibnamefont {Kumar}}, \
  and\ \bibinfo {author} {\bibfnamefont {A.}~\bibnamefont {Sundaresan}},\
  }\bibfield  {title} {\enquote {\bibinfo {title} {Surface ferromagnetism and
  superconducting properties of nanocrystalline niobium nitride},}\ }\href
  {\doibase https://doi.org/10.1016/j.matchemphys.2013.01.048} {\bibfield
  {journal} {\bibinfo  {journal} {Materials Chemistry and Physics}\ }\textbf
  {\bibinfo {volume} {139}},\ \bibinfo {pages} {500--505} (\bibinfo {year}
  {2013})}\BibitemShut {NoStop}%
\bibitem [{\citenamefont {Urbano}\ \emph {et~al.}(2002)\citenamefont {Urbano},
  \citenamefont {Pagliuso}, \citenamefont {Rettori}, \citenamefont
  {Kopelevich}, \citenamefont {Moreno},\ and\ \citenamefont
  {Sarrao}}]{Urbano2002}%
  \BibitemOpen
  \bibfield  {author} {\bibinfo {author} {\bibfnamefont {R.~R.}\ \bibnamefont
  {Urbano}}, \bibinfo {author} {\bibfnamefont {P.~G.}\ \bibnamefont
  {Pagliuso}}, \bibinfo {author} {\bibfnamefont {C.}~\bibnamefont {Rettori}},
  \bibinfo {author} {\bibfnamefont {Y.}~\bibnamefont {Kopelevich}}, \bibinfo
  {author} {\bibfnamefont {N.~O.}\ \bibnamefont {Moreno}}, \ and\ \bibinfo
  {author} {\bibfnamefont {J.~L.}\ \bibnamefont {Sarrao}},\ }\bibfield  {title}
  {\enquote {\bibinfo {title} {Field distribution and flux-line depinning in
  $\mathrm{M}\mathrm{g}{\mathrm{b}}_{2}$},}\ }\href {\doibase
  10.1103/PhysRevLett.89.087602} {\bibfield  {journal} {\bibinfo  {journal}
  {Phys. Rev. Lett.}\ }\textbf {\bibinfo {volume} {89}},\ \bibinfo {pages}
  {087602} (\bibinfo {year} {2002})}\BibitemShut {NoStop}%
\bibitem [{\citenamefont {Holm}\ and\ \citenamefont
  {Meissner}(1932)}]{holm1932}%
  \BibitemOpen
  \bibfield  {author} {\bibinfo {author} {\bibfnamefont {R.}~\bibnamefont
  {Holm}}\ and\ \bibinfo {author} {\bibfnamefont {W.}~\bibnamefont
  {Meissner}},\ }\bibfield  {title} {\enquote {\bibinfo {title} {Messungen mit
  hilfe von fl{\"u}ssigem helium. xiii: Kontaktwiderstand zwischen supraleitern
  und nichtsupraleitern},}\ }\href@noop {} {\bibfield  {journal} {\bibinfo
  {journal} {Zeitschrift f{\"u}r Physik}\ }\textbf {\bibinfo {volume} {74}},\
  \bibinfo {pages} {715--735} (\bibinfo {year} {1932})}\BibitemShut {NoStop}%
\end{thebibliography}%
\newpage

\end{document}